\documentclass[reprint,amsmath,amssymb,aps,pra,notitlepage,twocolumn]{revtex4-2}
\usepackage[OT4]{fontenc}
\usepackage{graphicx}
\usepackage{dcolumn}
\usepackage{bm}
\usepackage{color}

\begin{document}


\title{Self-consistent Description of Bose-Bose Droplets:\\ 
Harmonically Trapped Quasi-2D Droplets}



\author{Pawe{\l} Zin}
\affiliation{National Centre for Nuclear Research, ul. Pasteura 7, PL-02-093 Warsaw, Poland}

\author{Maciej Pylak}
\affiliation{National Centre for Nuclear Research, ul. Pasteura 7, PL-02-093 Warsaw, Poland}


\author{Mariusz Gajda}
\affiliation{Institute of Physics, Polish Academy of Sciences, Aleja Lotnik\'ow 32/46, PL-02-668 Warsaw, Poland}





\newcommand{\x}{{\bf r}}
\newcommand{\K}{{\bf k}}
\newcommand{\dk}{  \Delta {\bf k}}
\newcommand{\DK}{\Delta {\bf K}}
\newcommand{\KK}{{\bf K}}
\newcommand{\X}{{\bf R}}

\newcommand{\B}[1]{\mathbf{#1}} 
\newcommand{\f}[1]{\textrm{#1}} 

\newcommand{\half}{{\frac{1}{2}}}

\newcommand{\vv}{{\bf v}}
\newcommand{\p}{{\bf p}}

\newcommand{\dx}{\Delta {\bf r}}

\begin{abstract}
We describe a quantum droplet of a  Bose-Bose mixture squeezed by an external harmonic forces in one spatial direction. Our approach is based on the self-consistent method formulated in \cite{Pierwsza_praca}. The true spatial droplet profile in the direction of confinement  is accounted for, however local density approximation is assumed in the free directions. We define a numerical approach to find  the beyond-mean-field contribution to the chemical potential (Lee-Huang-Yang chemical potential) -- the quantity that  determines the droplet's profile. In addition to the numerical approach, we find  the Lee-Huang-Yang potential in the analytic form in two  limiting cases: a perturbative result for a strong confinement and a semiclassical expression when confinement is very weak.

\end{abstract}

\maketitle

\section{Introduction}

Problem of finding the ground state energy  of  weakly interacting, homogeneous Bose gas belongs to classic issues of quantum many body theory. It  attracts attention  of researchers 
\cite{VanHove52,Bethe56,Hugenholtz57} since  discovery of superfluidity of liquid Helium.  Huang and Yang  \cite{Huang56} studying energy of N-particle quantum system of hard-spheres and exploring  a concept of  Fermi pseudo-potential \cite{Fermi36} have shown that the ground state energy depends on the s-wave phase shift, proportional to the s-wave scattering length $a$. This observation allows to use as theoretical tools some other potentials, for instance smooth and positive defined,  instead of realistic van der Waals ones, provided that the scattering length is the same for both scattering potentials. Next to leading term in the expansion of energy in powers of density is known as the Lee-Huang-Yang (LHY) energy, \cite{LHY}. The energy of homogeneous weakly interacting Bose gas can be approximated by:
\begin{equation}\label{GSEF}
    E_0/N +E_{LHY}/N=\frac{2\pi a \hbar^2}{m_a} n \left( 1+\frac{128}{15\sqrt{\pi}}\sqrt{na^3}\right).
\end{equation}
In the above formula $E_0/N$ and $E_{LHY}/N$ are mean-field and LHY energy per atom, $a$ is the s-wave scattering length, $m_a$ is the mass of the atom and $n$ is atomic density. Further investigations of exited states proved  phonon-like momentum-energy spectrum at low energies \cite{Bogoliubov,Brueckner57,Girardeau59,Takano61}, and lack of energy gap  separating  ground and  exited states. This  feature was proved in \cite{Hugenholtz59}, and is referred to as the Hugenholtz-Pines theorem. Ground state energy of one and two dimensional  systems are also considered, \cite{Schick71, Popov72}, including exact N-particle solution of the Lieb-Linniger model \cite{Lieb63}. 

Advent of experiments with ultracold atomic gases leading to achievement of Bose-Einstein condensation  renewed interest in studies of ground state energy of  these dilute, weakly interacting systems \cite{Weiss04,Derezinski09}. 
 The formula giving the mean-field energy per atom 
$E_0/N$ present in
Eq.~(\ref{GSEF}) was proved with mathematical rigorousness \cite{Lieb98}  only in 1998 by E.H.~Lieb and J.~Yngvason who commented on difficulties which are pertinent to this issue: {\it "Owing to the delicate and peculiar nature of bosonic correlations, four decades of research have failed to establish this plausible formula rigorously."} Nowadays not only 3D but also lower dimensional systems are available to experiments with ultracold atoms. Theory follows this experimental progress.  First correction beyond Bogoliubov theory in the case of the ground state of two-dimensional weakly interacting Bose gas  was derived analytically in a systematic way \cite{Mora09} in excellent agreement with numerical Monte-Carlo calculations \cite{Pilati05,Astrakharchik09}.  

In most of experimental arrangements, the leading term in expression giving the ground state energy is sufficient to describe observations. The reason is that Bose condensates are very dilute and observation of subtle effects of the LHY contribution is beyond experimental precision.  Scientists used to share a folk wisdom that  higher order effects are to be observed at strong interactions.  This way of reasoning  was challenged by D.~Petrov \cite{Petrov15}, who noticed that one should rather look for such situations where leading term is small or vanishes. In such a case the higher order corrections will be dominating at low densities, thereby in the case of weakly interacting systems.

To make a men-field energy negligible one needs a system where attractive and repulsive interaction coexist and nearly cancel each other. Such conditions can be met in a two-component mixture when intraspecies scattering lengths are positive, $a_{11}, a_{22}>0$ but interspecies scattering length is negative, $a_{12}<0$, \cite{Petrov15}. Here $i=1,2$ enumerates the components. 

Alternatively one might consider one component systems where dipole-dipole interactions equalize zero-range repulsion. Such situation can be arranged in a prolate geometry when head-to-tail orientation of magnetic moments of atoms ensures effective attraction of dipoles \cite{Goral}. 

Indeed, while decreasing strengths of repulsive contact interactions using a Feshbach resonance, instead of expected collapse of atomic cloud, the Rosensweig instability leading to formation of an array of self-bound droplets was observed \cite{Kadau15}. 
This experiment triggered intensive experimental and theoretical studies of quantum droplets. They can be formed not only in dipolar systems \cite{Ferrier16a,Schmitt16,Ferrier16b,Chomaz16} but also in a mixture of two bosonic species \cite{Cabrera18,Cheiney18,Semeghini18,Derrico19}. 

Mechanism of formation of droplets is  described in \cite{Petrov15}. When  effective attraction overcomes repulsion, a collapse of the system (increasing of its density) is expected.  The mean-field energy is very small and negative, therefore the LHY term comes into a play. Collapse is arrested because of stabilizing, repulsive character of the higher order contribution to the energy. 

Formation of quantum droplets is  a very spectacular manifestation  of  higher order corrections to the system  energy. Precise knowledge of these contributions  is  crucial for understanding of droplet's properties.  
The LHY contribution originates in quantum fluctuations of Bogoliubov vacuum. It strongly depends on a phase space density, therefore dimensionality of the system matters. Contribution of quantum fluctuations to the mean field-energy of two-component Bose-Bose mixtures in reduced dimensions \cite{Petrov16} as well as at dimensional crossover \cite{Zin18,Ilg18} are one of the central issues of theoretical investigations. Similarly, the LHY energy of a one-component Bose gas with dipole-dipole interactions in  1D and 2D  geometries \cite{Edler17,Jachymski18} and at 2D-3D crossover \cite{Zin21} was found.  

Most of  theoretical results concerning systems in  lower dimensions  assume uniform density. However,  low dimensional configurations are reached by squeezing atomic clouds in one or two directions by harmonic potentials. Density profiles are not uniform thus.  Effect of external optical lattice potential is discussed in \cite{Kumlin19} but only situations where homogeneous approximation is justified are considered. 

The only to date result which accounts for a true density profile is given in \cite{Ilg18} where the LHY energy term of a droplet  squeezed by 1D or 2D harmonic potential is given. In quasi-2D exclusively  strong confinement case is studied.   The authors assume an universal regime, where a ratio of the mean-field energy to the confining potential excitation is the single parameter of the system. While calculating  Bogoliubov modes, responsible for  stabilization of  droplet, the local density approximations is used in  unconfined directions. Uniform lower-dimensional density, $n_{2d}$ or $n_{1d}$, depending on a number of free directions is assumed then.
The approach allows to find  the LHY energy (or alternatively LHY chemical potential) being the function of the lower-dimensional density. This LHY energy becomes an important contribution to the effective lower-dimensional  energy  functional, which enables to find droplets density profile in the unconfined direction.

This paper is, to some extend, continuation of studies presented in \cite{Ilg18}.  We find contribution to the chemical potential originating in quantum fluctuations for a  Bose-Bose droplet  squeezed by a harmonic potential in one spatial dimension,   while not confined in  two remaining dimensions. We consider  geometrical settings  in the {\it entire range of values of  aspect ratio, from two to three dimensions} -- at the whole crossover regime,   the most interesting from experimental point of view. 

Modified Gapless Hartree-Fock Bogoliubov method (MGHFB), introduced by us in \cite{Pierwsza_praca},  allows  to find the LHY energy at the crossover, not assuming uniform density profile {\it in confined direction}.  The method relies on coupled Generalized Gross-Pitaevski equation (GGP) and Bogoliubov-de Gennes equations. GGP  equation  accounts for energy related to quantum depletion and anomalous (regularized) density. 

We pay a special efforts to assure  gapless phonon-like spectrum of excitation. This is highly nontrivial issue, as the most natural attempt to improve over the Bogoliubov approach fails with this respect and contradicts Hugenholtz-Pines theorem. Accurate description of low energy physics is crucial to get right values of  beyond mean-field energy. 
The issue is extensively discussed in \cite{Morgan} in a context of  single component Bose system at temperatures close to a critical one and also in \cite{Derezinski09}.

In our studies we focus on a symmetric mixture
i.e. $g = g_{11}=g_{22}$, $\delta g = g_{12}+g$,
$N = N_1=N_2$ where $g = \frac{4\pi \hbar^2 a}{m_a}$ and so on.
We present a method which enables to calculate droplet's density profile.
It depends on four parameters $g$, $\delta g$, $N$, and harmonic confinement of frequency $\omega_z$. This is too many for a general treatment. Therefore we focus on a 
special case. 

We assume that  mean field interaction energy $\delta g  n$ (here $n$ is a peak atomic density) as well as contribution to the total energy originating in quantum fluctuations, $g \delta n$, $g m^R$, are much smaller than harmonic excitation energy $\hbar \omega_z$ ($\delta n$ and $m^R$ are quantum depletion and regularized anomalous density, respectively).  
In such case deviations of droplet's density profile from  a density of ground state of  harmonic confinement is negligible. Similarly as in \cite{Ilg18} we assume uniform $n_{2d}$ density and  use the local density approximation in  solutions of  Bogoliubov equations.  Effectively we solve a one-dimensional problem, though. 

In the regime described above (which we call universal regime), the three (dimensionless) parameters which control the system, reduce to a single one, $ y = \frac{g n_{2d}}{a_{ho} \hbar \omega} \approx \frac{g n}{\hbar \omega_z}$, where  2D atomic density, $n_{2d}$, is of the order of $n_{2d} \approx n a_{ho}$, $a_{ho}$ being harmonic oscillator length. 
We shall  mention that mean-field energy, $gn$, can be much larger than $\hbar \omega_z$. As a consequence $y$ can be much larger than unity.

Our paper is organized as follows. In Sec.~\ref{sec2} we briefly describe MGHFB method.
In Sec.~\ref{harm2d}, based on MGHFB approach, we find quantum-fluctuation-contribution (LHY chemical potential) to the chemical potential of the system $\mu_{LHY}(y)$ as a function of $y$. In the quasi-2D limit, i.e. if $y \ll 1$, a perturbative approach can be used and analytical formula for  $\mu_{LHY}^{pert}$ can be obtained. On the other hand if $y \gg 1$ we expect to recover the 3D analytical result obtained using local density approximation in expression for uniform system. Indeed our numerical calculation agree with analytical formulas in both limits. In addition, the numerical result is given in the entire range of quasi-2D to 3D  transition. Summary and final conclusions are presented in Sec.~\ref{harm2d}. Lengthy calculations are moved to Appendixes. In particular in Appendix~\ref{app:har} we show semiclassical results in limit of $y > 1$, while in Appendix~\ref{har2} perturbative calculations in the case of $y\ll 1$ are presented.

\section{Bose-Bose mixture }  \label{sec2}
Detailed derivation of equations of  the Modified Gapless Hartree-Fock Bogoliubov method (MGHFB) of description of  a Bose-Bose mixture is presented in \cite{Pierwsza_praca}. Here  we give a summary of major ideas.
Interaction Hamiltonian of a Bose-Bose mixture   involves three interaction potentials, $U_{11}, U_{22}$ and $U_{12}$:
\begin{eqnarray}
&&    H_{int}=\frac{1}{2}\sum_{i=1,2}\int \mbox{d}\x  \mbox{d}\x' \hat \psi_i^\dagger(\x) \hat \psi_i^\dagger (\x')U_{ii}(\x-\x')\hat \psi_i(\x) \hat \psi_i(\x') \nonumber \\
&&    +\int \mbox{d}\x \, \mbox{d}\x' \hat \psi_1^\dagger(\x) \hat \psi_2^\dagger(\x')U_{12}(\x-\x')\hat \psi_1(\x) \hat \psi_2(\x'),
\end{eqnarray}
where  $\hat \psi_i(\x)$ ($i=1,2$) are field operators of the two droplet's components. Total Hamiltonian includes also a single particle contributions $H_0(\x)$ being a sum of the kinetic and potential energy terms, $H_{sp}=\sum_i \int \mbox{d}\x \hat \psi_i^\dagger(\x) \hat H_0(\x) \hat \psi_i(\x)$.

At low scattering energies the standard mean field approach is based on the assumption that  $U_{ij}(\x)$ can be approximated by a contact potentials  $U_{ij}(\x) \approx g_{ij} \delta(\x)$ where interaction strengths $g_{ij}=\frac{4 \pi \hbar^2}{m_a}a_{ij}$ are proportional to the s-wave scattering lengths,  $a_{ij}$, of the interaction potentials. In the following we  focus on a case when  inter-scatterings lengths $a_{ii}$ are positive (effective interaction is repulsive) and $a_{12}$ is negative (effective interaction is attractive). In addition, to simplify calculations, we assume that the two components have equal masses $m$. Mean values of the field operators are assumed to be different than zero because both species  are  Bose-condensed. Accordingly, we split these operators to explicitly distinguish the mean field, $\psi_i = \langle \hat \psi_i \rangle$, and small quantum perturbations, $\hat \delta_i$, \cite{stringari}: 
\begin{equation}
\label{hatpsi}
\hat \psi_i = \psi_i +\hat \delta_i,
\end{equation}
The Hamilton equations lead to the following set of the two coupled stationary GP  equations:
\begin{eqnarray}
\label{SGP1}
&& \mu_1 \psi_1(\x) = H_0 \psi_1 + g_{11} |\psi_1|^2 \psi_1 + 2 g_{12} |\psi_2|^2 \psi_1,
\\
\label{SGP2}
&& \mu_2 \psi_2(\x) = H_0 \psi_2 + g_{22} |\psi_2|^2 \psi_2 + 2 g_{12} |\psi_1|^2 \psi_2,
\end{eqnarray}
if  quantum fluctuations are neglected.
In the above $\mu_i$ are chemical potentials of the species and mean fields are normalized to the total number of atoms of each kind $\int \mbox{d}\x |\psi_i|^2 = N_i$. 
It is quite intuitive that if attractive interaction is weak the gas would fill the whole space allowed by a confining potential. If  attractive interaction grows the instability appears  at a certain critical value. After crossing this point the gas tends to increase the density. The above mean field approach predicts  a transition  from a stationary solution to a state which  eventually collapses (tends to infinite density).  In a simplest case of an uniform mixture of species with equal masses the instability occurs when $\sqrt{g_{11}g_{22}}+g_{12} \leq 0$.

D.~Petrov \cite{Petrov15} noticed that at transition point, $\sqrt{g_{11}g_{22}}+g_{12} = 0$, where  mean field energy vanishes, the higher order contributions to the system energy   must be accounted for  in  Eqs.~(\ref{SGP1}) and (\ref{SGP2}). These terms originate in quantum fluctuations and are responsible for stopping a collapse and formation of droplets. 

In the following, we assume  a symmetric situation, i.e. the same interparticle interaction potential for both species, $U_{11}(\x) = U_{22}(\x) = U(\x)$,  and the same number of atoms. Obviously  both mean fields are equal then, $\psi_1(\x) = \psi_2(\x)$.
Excitations can be divided into  soft and hard modes which in  a symmetric case are:
\begin{eqnarray}
\label{mfmodes}
\hat \psi_\pm   & = & \frac{1}{\sqrt{2}} \left( \hat \psi_1 \pm \hat \psi_2 \right),\\
\hat \delta_\pm & = & \frac{1}{\sqrt{2}} \left( \hat \delta_1 \pm \hat \delta_2 \right). 
\end{eqnarray} 
Introducing similar combinations of the mean-fields, $\psi_+ = \psi_1+ \psi_2 $ and $\psi_-= \psi_1- \psi_2 =0$ and  using Eq.~(\ref{hatpsi})  we find that:
\begin{eqnarray}
\label{deltamodes}
\hat \psi_+ & = & \psi_+ + \hat \delta_+ , \\
\hat \psi_- & = & \hat \delta_-.
\end{eqnarray}
The $\psi_+(\x)$-component  corresponds to the soft-mode mean-field, and $\psi_-(\x) = 0$ is the mean-field of the hard mode, equal to zero in the symmetric case. It follows from Bogoliubov equations \cite{Petrov15,sacha} that fluctuations breaking the symmetry between the species, i.e. described by $\hat \delta_{-} (\x)$, are energetically very costly. These  hard mode excitations are characterized by a large sound velocity giving a large contribution to the energy of quantum fluctuations. On the contrary, excitations of the soft mode, $\hat \delta_+(\x)$, are characterized by a small sound velocity.
Their impact on the LHY energy, close to the critical point, is small.
As we shall work close to the critical point we neglect the contribution of the soft modes to the LHY energy.


To simplify notations we define $\psi(\x) \equiv \psi_+(\x)$ and $\hat \delta(\x) \equiv  \hat \delta_-(\x)$. We additionally assume $\psi$ o be a real function.
This way instead of the two mean fields $\psi_i(\x)$ and two quantum fields $\hat \delta_i(\x)$ we consider only one mean field - the soft mode mean-field, and one quantum-fluctuation operator -- hard mode fluctuations. The problem is simplified thus to a single component Bose field $\hat \psi(\x)$ having $2N$ atoms in total. 

Generalized Gross Pitaevskii equation accounting for quantum fluctuations of the hard mode takes the form:
\begin{widetext}
\begin{eqnarray} \label{GP}
&& 0 = (H_0 - \mu) \psi(\x) + 
\int \mbox{d} \x' \,  U_s(\x-\x')
\left(  \psi^2(\x') + \delta n(\x',\x') \right) \psi(\x)
\\ \nonumber
&& 
 + \int \mbox{d} \x' \, U_d(\x-\x')
\left( \delta n(\x',\x)  \psi(\x') + m(\x',\x)  \psi(\x') \right)
\end{eqnarray}
\end{widetext}
where $\delta n(\x',\x) =  \langle \hat \delta^\dagger(\x',t)  \hat \delta(\x,t) \rangle $
, $m(\x',\x) = \langle \hat \delta(\x',t)  \hat \delta(\x,t) \rangle $,
$U_s = (U+U_{12})/2$ and $ U_d = (U-U_{12})/2$. Quantum contributions  $\delta n$ and $m$ are real functions do not depending on time.

Accounting for quantum fluctuations forces us to treat with a special care both low and high energy components of the interaction potentials $U_{ij}$.  In the following we assume that all potentials $U_{ij}(\x)$ have a bell-like shape of a characteristic widths $\sigma_{ij}$ respectively, and all widths are of the same order, $\sigma_{ij} \sim \sigma$, being much
larger than the all s-wave scattering lengths, $ |a_{ij}| \sim a$ and much smaller than two other length scales, i.e.: \\
i) the healing length $ \xi(\x)=\frac{\hbar}{\sqrt{m_a n |\delta g| }}$, a quantity determining a radial size of a droplet's surface, where $n(\x)$ is atomic density, $\delta g = \sqrt{g_{11}g_{22}}+g_{12}<0$  and 
$g_{ij}=\frac{4\pi \hbar^2 a_{ij}}{m_a}$,\\
ii) Characteristic distance $d$ of density variations. The nonuniform density profile results from squeezing of a droplet in a direction of external potential.

If $a \ll \sigma \ll \xi, d $,  all properties of a droplet depend exclusively on  low energy scattering properties of the interaction potentials, namely the $s$-waves scattering lengths $a_{ij}$. Therefore potentials $U_{ij}$ can be approximated by their lowest order Fourier components $\tilde u_{ij}(0)$ which, in turn,  can be related to the $T$-matrix expansion of the  scattering potentials, $g_{ij} = \tilde u_{ij}(0)  - \frac{1}{(2\pi)^3} \int \mbox{d} \K \, \frac{\tilde u_{ij}^2(\K)}{2 E_k}$, where $E_k=\frac{\hbar^2 k^2}{2m_a}$.  Whenever $\tilde u_{ij}(0)$ multiplies a small quantity, like $\delta n_i$ i.e. fluctuations of density of atoms,  it is sufficient to approximate $\tilde u_{ij}(0) \approx g_{ij}$. This is equivalent to substitution $U_{ij}(\x) = g_{ij} \delta(\x)$. However if $\tilde u_{ij}(0)$ multiplies condensate density in a given component, $n_i$,  we shall keep also the second order term in the Born expansion  of the $T$-matrix.

According to the above discussion, the GGPE depends only on the low energy scattering properties of the interaction potentials:
\begin{equation} \label{finalGP}
0 = \left(H_0 - \mu + \frac{\delta g}{2}  \psi^2(\x)   +  g \delta n(\x) + g m^R(\x) \right)   \psi(\x), 
\end{equation}
where the normalization condition is:
\begin{equation}
    \int d \x \left( \psi^2(\x) + \delta n(\x) \right) = 2N.
\end{equation}
In Eq.(\ref{finalGP})  we introduced $g  = - g_{12} +\delta g $ and assumed $\psi(\x,t) = e^{- i \mu t} \psi(\x) $. 

The second order terms depending on  high energy modes  conspire together with the  anomalous density to give a regularized anomalous density (assuming $U_{12}(\x) \simeq - U(\x)$):
\begin{eqnarray} \nonumber
g m^R(\x)  &=&  \int  \frac{\mbox{d} \K}{(2\pi)^3}\, \frac{\tilde u^2(\K)}{2 E_k}   \psi^2(\x) + \int \mbox{d} \x' \, U(\x-\x')  m(\x',\x),\\
\label{mRinne}
\end{eqnarray}
which depends  only on low momenta $k \simeq 0$ part of the interaction potentials i.e. on their scattering length  only.

To get stationary Bogoliubov equations we factorize a time dependence of fluctuation operator and use  standard expansion into   eigenmodes  $ u_{\nu}(\x), u_{\nu}(\x)$, i.e.
$\hat \delta(\x,t) = e^{-i \mu t} \left( \sum_\nu u_{\nu}(\x) e^{- i \varepsilon_{\nu} t} 
\hat \alpha_{\nu}
+ v_{\nu}^*(\x)  e^{ i \varepsilon_{\nu} t} \hat \alpha_{\nu}^\dagger \right)$.
Combining this expansion with linear Heisenberg equations for $\hat \delta(\x,t)$ gives:
\begin{eqnarray}\nonumber
&&  \left(H_0-\mu_0+   g \psi^2(\x) \right)   u_{\nu}(\x) +   g \psi^2(\x)   v_{\nu}(\x)  = \varepsilon_{\nu}   u_{\nu}(\x) 
\\ \label{Bog-2}
\\ \nonumber
&& \left( H_0-\mu_0+   g \psi^2(\x) \right)   v_{\nu}(\x)  
+    g\psi^2(\x)  u_{\nu}(\x) = - \varepsilon_{\nu} v_{\nu}(\x).
\end{eqnarray}
Note that in Eq.(\ref{Bog-2}) the chemical potential $\mu$ is substituted by $\mu_0$.
This is a crucial element of MGHFB method.
The chemical potential $\mu_0$ has  to be found from  Bogoliubov equation determining  the zero-mode wavefunction, \cite{Zin21}, $u_{0}(\x) = - v_{0}(\x)$: 
\begin{equation}\label{Eqmu0}
(H_0 - \mu_0) u_{0}(\x) = 0.
\end{equation}
Note, that  excitation energy is set to zero, $\varepsilon_{0} =0$ in Eq.~(\ref{Eqmu0}).
As discussed in details in \cite{Pierwsza_praca} the replacement $\mu \to \mu_0$ is necessary to get a consistent gapless approach  and phononic branch in the excitation spectrum. It ensures that  amplitudes of Bogoliubov modes have a correct limit at low energies. 
Substitution of $\mu$ by $\mu_0$ is justified because
$|\mu-\mu_0|$ is much smaller than characteristic interaction
energy term, $gn$, which enters  Bogoliubov equations
(\ref{Bog-2}).
The replacement is consistent with other approximations, though.

Solutions of  the Bogoliubov equations allow to find  quantum depletion:
\begin{equation}
    \delta n(\x) = \sum_{\nu \neq 0} |v_{\nu}(\x)|^2,
\end{equation} 
and renormalized anomalous density:
\begin{equation}
m^R(\x) = \frac{\partial}{ \partial |\dx|} \left( |\dx| m \left(\X  ,\dx \right)   \right)_{\dx \rightarrow 0},    
\end{equation}
where 
\begin{equation}
    m(\x,\x') = \sum_{\nu \neq 0} u_\nu(\x) v_\nu^*(\x'),
\end{equation}
and  $\X = \frac{\x+\x'}{2}$, and $\dx = \x-\x'$.

We now consider a system where numerical solution of the Bogoliubov equations (\ref{Bog-2}) is necessary. In such case quantum fluctuation terms are to be split into low and high energy parts: $\delta n = \delta n_L + \delta n_H$ and $m_R = m_L +m^R_H$. Low energy components are to be found numerically directly from definitions while the high energy components can be obtained using semiclassical approximation:
\begin{eqnarray} \nonumber 
&& m_H^R(\x) = g  \psi^2\left(\x \right)  \frac{m}{2\pi^2 \hbar^2} k_c(\X) 
\\ \label{mRnMieszanina}
&& 
+ 
g \psi^2(\x) \int \mbox{d} \Omega_\K  \int_{k_c(\X)}^\infty 
\frac{ k^2 \mbox{d} k }{(2\pi)^3}
\left( \frac{1}{\frac{\hbar^2 k^2}{m_a}}  -  \frac{1}{2 \varepsilon(k,\X)  }  \right)   
\end{eqnarray}
where
\begin{eqnarray}
\varepsilon(k,\x)= \sqrt{ \left( A(\K,\x) \right)^2 -\left(   g \psi^2(\x)  \right)^2  },
\end{eqnarray} 
where  $k_c(\x)$ is given by equation $\varepsilon_{\nu}(k_c(\x),\x) = E_c  $ and 
$A(\K,\x) = \frac{\hbar^2 k^2}{2m_a}  + V(\x) - \mu_0   + g   \psi^2(\x) $. Using the same method we find
\begin{eqnarray}
 \delta n_{H}(\x) \simeq    \int \mbox{d} \Omega_\K  \int_{k_c(\x)}^\infty 
\frac{ k^2 \mbox{d} k }{(2\pi)^3} 
\frac{1}{2} \left( \frac{A(k,\x)}{  \varepsilon(k,\x) }  -1   \right),
\end{eqnarray}
Note, that the problem has to be solved self-consistently because $\delta n$ and $m^R$ depend on $\psi$ which in turn is a solution of GGPE, Eq.~(\ref{finalGP}), which involves  $\delta n$ and $m^R$ as essential ingredients. The above equations define the self consistent method that enables to determine  droplet's wave function $\psi(\x)$.
\\

\section{Harmonically confined quasi-2d Bose-Bose system} \label{harm2d}

\subsection{Universal regime}

We now move to physically important case of a Bose-Bose mixture confined in one  spatial direction (we choose it to be the $z$-direction) by a harmonic potential $V(z) = \frac{1}{2} m_a \omega_z^2 z^2$.
The system has a `pancake' geometry and $\x_\perp = x {\bf e}_x + y {\bf e}_y$ is a vector
in a plane perpendicular to $z$-axis.
Such a system was analyzed in \cite{Petrov16} in a  strong confinement limit. Here we want  to describe 
the system in the entire range of possible arrangements,  from quasi-2D  to 3D geometry what can be  achieved  by changing  strength of the  confinement. 

To find the LHY contribution to a chemical potential  we shall use numerical solutions of  Bogoliubov equations to obtain $\delta n$ and $m^R$ at the transition point.  We additionally restrict  our considerations to such arrangements  for which  the excitation energy in a tight direction is much larger than `low energies of the problem':
$ \frac{\delta g}{2}  \psi^2(\x) , g \delta n(\x) ,  g m^R(\x)  \ll \hbar \omega_z$. 
Note, however that we do not assume that $\hbar \omega_z$ must be larger than $ g \psi^2(\x)$, therefore our considerations include also 3D case.

Under the above conditions  a solution of Eq.~(\ref{finalGP}) is well approximated by  
\begin{equation}\label{ansatz}
\psi(\x) = \psi_\perp(\x_\perp) \phi_0(z)  
\end{equation}
where $\phi_0(z)$ is normalized to unity ground state of the harmonic oscillator,  $\int d z \,  \phi_0^2(z) =1 $. 
GGPE following from Eq.(\ref{finalGP}) is   
\begin{widetext}
\begin{equation}\label{muN1}
\left( - \frac{\hbar^2}{2m} \triangle_\perp -\mu + \frac{1}{2} \hbar \omega_z + \frac{\delta g}{2} \psi_\perp^2(\x_\perp) 
\int \mbox{d}z \, \phi_0^4(z)
+  g  \int \mbox{d} z \, \left(\delta n(\x_\perp,z) + m^R(\x_\perp,z)  \right) \phi_0^2(z) \right) \psi_\perp(\x_\perp) = 0.
\end{equation}
\end{widetext}

A short comment on a validity of the ansatz $ \psi(\x) \simeq  \psi_\perp(\x_\perp) \phi_0(z) $  is now in order. In fact, assumption that $z$-dependence of the mean-field wavefunction $\psi(\x)$ is the same as those of harmonic oscillator $\phi_0(z)$ is approximate.  The terms  $ \frac{\delta g}{2}  \psi^2(\x) , g \delta n(\x) ,  g m^R(\x) $ in Eq.~(\ref{finalGP})  introduce some deviations of $\phi_0(z)$ from the ground state of harmonic oscillator. The `back-action', i.e. an effect of modification the quantities of interest,  $\delta n(\x)$ and $m^R(\x)$,
by a `disturbed' $\phi_0(z)$  is negligible because they do not appear  explicitly in the  Bogoliubov equations, Eq.~(\ref{Bog-2}). 
Neither $\mu_0$ depends on them,  Eq.~(\ref{Eqmu0}).  
Here we do not take into account this small modification.

\subsection{Local density approximation in   $\x_\perp$-direction}

Eq.(\ref{muN1}) is written in natural units in order to give a clear physical picture of  individual terms. Here we  switch to  harmonic oscillator unit of distance $a_{ho}=\sqrt{\frac{\hbar}{m_a\omega_z}}$ and  energy $\hbar \omega_z$.
Therefore, from now on,  wavefunctions  $\phi_0(z)$ and $\psi_\perp(\x_\perp)$ as well as quantum depletion and anomalous (renormalized) density are  dimensionless.  We do not introduce new notation for dimensionless quantities, however.

We now notice  that characteristic length scale associated with   changes of  droplet's density in  free directions, $\x_\perp$, roughly equals to $\xi = \hbar/\sqrt{ m_a |\delta g| n}$,
and is much larger than $a_{ho}$ (this condition follows from assumption that
 $|\delta g| n \ll \hbar \omega_z$). Therefore we can use local density approximation 
in $\x_\perp$ directions only. We solve Bogoliubov equations using this approximation.
\begin{widetext}
\begin{eqnarray} \nonumber
&& \left( \frac{k_\perp^2}{2} - \frac{1}{2} \partial_z^2 + \frac{1}{2} z^2 - \frac{1}{2}  + y \phi_0^2(z)  \right)u_{k_\perp,\nu}(y,z) 
+ y \phi_0^2(z) v_{k_\perp,\nu}(y,z) = \varepsilon_{k_\perp,\nu} u_{k_\perp,\nu}(y,z), 
\\  \label{BogHar}
&& \left( \frac{k_\perp^2}{2} - \frac{1}{2} \partial_z^2 + \frac{1}{2} z^2 - \frac{1}{2}  + y \phi_0^2(z)  \right)
v_{k_\perp,\nu}(y,z) + y \phi_0^2(z) u_{k_\perp,\nu}(y,z) =  - \varepsilon_{k_\perp,\nu} v_{k_\perp,\nu}(y,z),
\end{eqnarray}
\end{widetext}
where 
\begin{eqnarray}
&&u_{k_\perp,\nu}(\x) =    e^{ i \K_\perp \x_\perp}  u_{k_\perp,\nu}(z), \\
&&v_{k_\perp,\nu}(\x) =   e^{ i \K_\perp \x_\perp}  v_{k_\perp,\nu}(z)
\end{eqnarray}
and normalization condition is, $\int d z \,  |u_{k_\perp,\nu}(z)|^2 -  
|v_{k_\perp,\nu}(z)|^2 = 1.$
 The above are obtained from Eq.~(\ref{Bog-2})
where we substituted $-\triangle_\perp$ by $k_\perp^2$
and used the ansatz Eq.(\ref{ansatz}). 
$\x_\perp$ dependence is hidden in dimensionless parameter,  $y$,  given by the ratio of   mean-field energy per atom to  oscillator excitation energy:
\begin{equation}
\label{y}
  y = \frac{g \psi^2_\perp(\x_\perp) }{a_{ho}^3 \hbar \omega_z} = \frac{4 \pi a}{a_{ho}} \psi^2_{\perp}(\x_{\perp}).  
\end{equation}

Similarly,  Eq.~(\ref{muN1}) written in the oscillatory units is
\begin{eqnarray} \label{GP_har}
&&    \left( -\frac{1}{2} \triangle_\perp  + \frac{1}{2}
    +  y \frac{\delta g}{2g} \frac{1}{\sqrt{2\pi}} +  \Delta \mu(y) -\mu \right) \psi_\perp(\x_\perp)=0.\nonumber \\  
\end{eqnarray}
In  the above we used  $\phi_0^2(z) =  \pi^{-1/2} \exp (-z^2) $ and  $ \int \mbox{d}z \, \phi_0^4(z) = \frac{1}{\sqrt{2\pi}} $.   
From the ground state solution of Eq.~(\ref{Eqmu0}) we found  $\mu_0 = \frac{1}{2}$. 
A contribution to the chemical potential originating in quantum fluctuations (related to the Lee-Huang-Yang energy) is denoted by 
\begin{equation}\label{tilde_mu}
 \Delta \mu(y)=\frac{4\pi a}{a_{ho}}  \mu_{LHY}(y), \end{equation}
where $\mu_{LHY}$ is:
\begin{equation}
\label{mu_profile}
  \mu_{LHY}(y)  = 
 \int \mbox{d} z \, \left(     \delta n(y,z) + m^R(y,z)  \right) \phi_0^2(z),   
\end{equation}
The chemical potential Eq.(\ref{mu_profile}) is averaged with density profile $|\phi_0(z)|^2$ because we reduced 3D GGP equation to  2D form, Eq.(\ref{GP_har}), by integrating over  $z$-direction, assuming fixed harmonic oscillator $z$-component  wavefunction of a droplet.  

Formalism presented in this section defines a method of finding droplet's wave function $\psi_\perp(\x_\perp)$. First,  Bogoliubov equations are to be solved (\ref{BogHar}). The solutions allow to obtain  $\delta n(y,z)$ and $m^R(y,z)$ (in what follows we describe this calculation in more detail). Using Eq.~(\ref{mu_profile}) we calculate $\mu_{LHY}(y)$ for all values of $y$. 
Now, for  given  values of $a/a_{ho}$ and $\delta g/g$, all ingredients of the left hand side of  Eq.~(\ref{GP_har} ) are uniquely defined, and droplet's profile  $\psi_\perp(\x_\perp)$ and chemical potential $\mu$ can be found as the eigenstate and the  eigenenergy of the GGP equation. The wavefunction gives mean-filed of both species, thus   normalization condition reads:
\begin{equation}
\int d \x \, \left( \psi_\perp^2(\x_\perp) \phi_0^2(z) + \delta n(y,z) \right) = 2 N,     
\end{equation}
where $\psi_\perp$ enters  definition of  $y$. This completes  the method of determination of  $\psi_\perp(\x_\perp)$.

Finally, we stress that our result are restricted to 
systems being tightly confined in the $z$-direction:
$ \frac{\delta g}{2}  \psi^2(\x) , g \delta n(\x) ,  g m^R(\x)  \ll \hbar \omega_z$, 
what can be summarized by the following conditions:
\begin{eqnarray}
y \frac{\delta g}{2g} \frac{1}{\sqrt{2\pi}} \ll 1,
\\
 \Delta \mu(y) = \frac{4\pi a}{a_{ho}}  \mu_{LHY}(y)
\ll 1.
\end{eqnarray}
These conditions give  limits on a maximal strength of the interactions.

\subsection{Determination of $\mu_{LHY}(y)$}

To calculate  $\mu_{LHY}$ numerically, different   approaches  at low and high energies have to be used. To this end  quantum depletion 
and anomalous density are divided into low energy and high energy contributions, $\delta n = \delta n_{L}+\delta n_{H}$  and similarly $m = m_{L}+m_{H}$.
Low energy regime is defined by  conditions  $|\K_\perp| \leq k_c$ and $\varepsilon_{k_\perp,\nu} \leq \frac{k_\perp^2}{2} +  \frac{k_c^2}{2} $. 

In the case of nonuniform system when LDA cannot be used the  analytic calculations are much more complicated. At high energy sector the semiclassical method can be applied. It gives (for detailed calculations see Appendix \ref{app:har}):
\begin{eqnarray}\label{mH_sing}
&& m_{H}(y,z) \simeq  - y \phi_0^2(z) \frac{1}{4\pi |\dx|}
\\
&& + y \phi_0^2(z) k_c \frac{ \pi + \log 4 }{8 \pi^2}  h(z,y,k_c) 
+ m^R_{S,H}(y,z), \nonumber
\end{eqnarray}
where $h(z,y,k_c)$ and $m^R_{S,H}(y,z)$ is given by Eq.~(\ref{def:h}) and Eq.~(\ref{mRsemW}) respectively.
The main message of Eq.(\ref{mH_sing}) is to singled-out a singular term, $\propto \frac{1}{4\pi|\dx|}$, and obtain a regular but cut-off depend contribution:
\begin{equation}
\label{hmr}
 m^R_H(y,z) = m^R_{S,H}(y,z) +y \phi_0^2(z) k_c \frac{ \pi + \log 4 }{8 \pi^2}  h(z,y,k_c),   
\end{equation}
Regularized anomalous density, $m^R(y,z)$ involves both high and low energy sectors, $m^R(y,z)=  m^R_H(y,z)+ m_L(y,z)$, where the low-energy term, $m_L(z)$ is:
\begin{equation} \label{lmr}
m_L(y,z) = \frac{1}{2\pi} \int_0^{k_c} k_\perp \mbox{d} k_\perp 
\sum_{\nu \in {L}} u_{k_\perp,\nu}(y,z)   
v^*_{k_\perp,\nu}(y,z).
\end{equation}
"$L$" denotes a set of these  $\varepsilon_{k_\perp,\nu} $ which satisfy 
$\varepsilon_{k_\perp,\nu} \leq \frac{k_\perp^2}{2} +  \frac{k_c^2}{2} $.
Evidently the low energy contribution depends on the cut-off. 
Similarly, the high energy part of quantum depletion  $\delta n_{H}(z,y)$ is obtained using semiclassical method  in  Appendix \ref{app:har} and given by Eq.~(\ref{n_H}):
\begin{equation}\label{dnzP}
\delta n_H(y,z) =  \frac{\left( y \phi_0^2(z) \right)^{3/2}}{3 \pi^2} f(z,y,k_c).
\end{equation}
Low energy component can be found directly from the definition:
\begin{equation} \label{nLz}
\delta n_L(y,z) =  \frac{1}{2\pi} \int_0^{k_c} k_\perp \mbox{d} k_\perp \sum_{\nu \in {L}} 
\left|v_{k_\perp,\nu}(y,z)\right|^2.
\end{equation}
Finally, contribution of the Bogoliubov-vacuum fluctuations   to the chemical potential of the system,   after averaging over the $z$-direction density profile, Eq.(\ref{mu_profile}) is:
\begin{equation}
\label{mufin}
  \mu_{LHY} (y)= 
   \mu^H_{LHY}(y) +  \mu^L_{LHY}(y),
\end{equation}
where (see Eq.(\ref{delta_high})): 
\begin{eqnarray} \label{DHLHY}
&&  \mu^H_{LHY} (y) =  \frac{y^{3/2}}{3\pi^{11/4}}  \sqrt{\frac{2}{5}}\left(3 G(y,k_c)+   F(y,k_c) \right) \nonumber \\
&&  + y \, k_c  \frac{ \pi + \log 4  }{8 \sqrt{2} \pi^{5/2}} H(y,k_c),
\end{eqnarray}
and:
\begin{equation} \label{DLLHY}
 \mu^L_{LHY}(y,k_c) = \int {\rm d}z \phi_0^2(z) (\delta n_L(y,z)+\delta m_L(y,z)).
\end{equation}
Although, all individual terms in Eq.(\ref{mufin}) depend on the cut-off momentum, $k_c$, the whole expression does not. In general case $\mu_{LHY} (y)$ has to be evaluated numerically, however the high energy contribution is regularized  and expressed in a form of several integrals leading to smooth functions, $G(y,k_c)$, $F(y,k_c)$,  $H(y,k_c)$ specified in Appendix \ref{app:har}, see Eqs.~(\ref{mSHR}), (\ref{dnSHR}) and (\ref{HS}). 


This general formalism can be simplified in two  regimes. 
First, when the harmonic confinement is week, $y>1$ and the system has `almost continuous' spectrum, thus the semiclassical approximation, valid in principle for high momenta components, can be extended over the entire range of energies -- from zero up to a cut-off energy, which after renormalization of the anomalous density, can be sent to infinity. This procedure leads to the semiclassical expression for $ \mu^{semi}_{LHY}(y)$, see Appendix \ref{app:har}, Eqs.~(\ref{mu_semi}): 
\begin{equation}\label{mu_semi_main}
 \mu^{semi}_{LHY}(y) = 
 \frac{y^{3/2}}{3\pi^{11/4}} \sqrt{\frac{2}{5}}  \left( 3 G_{semi}(y) + F_{semi}(y) \right).
\end{equation}
\begin{figure}[htb]
    \centering
    \includegraphics[width=0.99\linewidth]{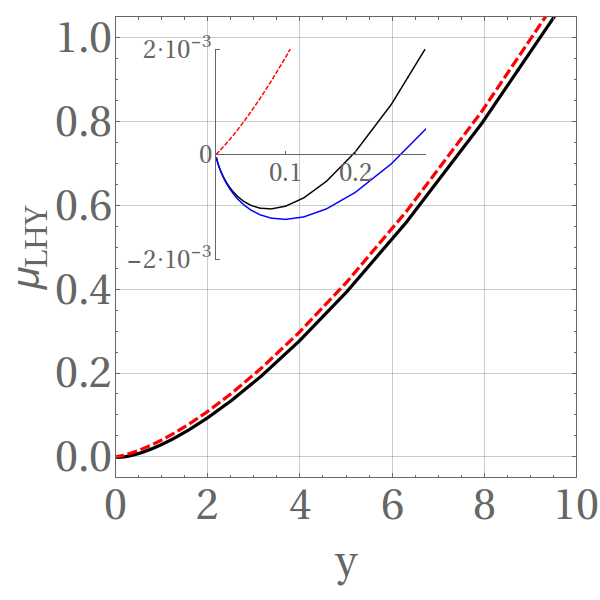}
    \caption{LHY contribution to the chemical potential, $\mu_{LHY}$, as a function of $y=\frac{g n_\perp}{a_{ho}\hbar\omega_z}$. Black line indicates numerical results while the red one semi-classical solution. Inset: black line  - numerical results, blue line - perturbative solution}
    \label{fig:dmiu1}
\end{figure}
On the other hand if $y \ll 1$,  perturbative calculations are possible. This is a situation  when the harmonic confinement is very tight and Bogoliubov amplitudes can be found in the lowest order perturbation of  eigenstates of harmonic oscillator. Details of these tedious calculations are presented in  Appendix~\ref{har2}. The final result  can be summarized as follows:
\begin{equation}\label{limit}
   \mu^{pert}_{LHY}(y) = 
\frac{y}{8\pi^2}   \log \left( \frac{y}{4\pi} C_{2d}^h \sqrt{e} \right),
\end{equation}
where $ C_{2d}^h \simeq 28.69 $. Eq.~(\ref{limit}) is in agreement with the result given in \cite{Petrov16}. This  is a very important test of our approach.

Contribution to the chemical potential originating in quantum fluctuations is plotted  in Fig.~(\ref{fig:dmiu1}). 
Here we plot $  \mu_{LHY}(y)$ as a function of $y$. Black line indicates numerical result given by Eq.~(\ref{mufin}).  Semiclassical result obtained with the help of Eq.(\ref{mu_semi_main}) is depicted  by the red line,  $  \mu^{semi}_{LHY}(y)$, Eq.~(\ref{limit}). 

In  inset of Fig.~(\ref{fig:dmiu1}) we  show results for $y \ll 1$,  where perturbative calculations are in order $ \mu_{LHY}(y) \simeq   \mu^{pert}_{LHY}(y)$, Eq.(\ref{limit}). This result  is plotted by the blue line. If $y\ll 1$ the LHY contribution to the chemical potential $ \mu_{LHY}$ is negative.  We notice that both black and blue curves are practically  identical if $y < 0.02$. The strong confinement  limit $y \ll 1$ is clearly visible in the numerical result, Eq.~(\ref{limit}).

Now we shortly discuss the semiclassical result Eq.~(\ref{mu_semi_main}).
Naively, it should agree with the LDA expression if $y \gg 1$, which  can be obtained by replacement of  atomic density, $n$, in the 3D  expression  describing a  uniform system,  $\Delta \mu_{3D} =  \frac{32}{3 \sqrt{\pi}} gn \sqrt{na^3}$, by a local  nonuniform density accounting in the case of 1D harmonic  confinement, $n=n_{\perp} \phi_0^2(z)$.  However, to use this expression in the quasi-2D formalism (2D GGP equation), it has to be integrated over $z$-coordinate with  1D density profile, $\phi_0^2(z) = \frac{1}{a_{ho}\sqrt{\pi}} \exp{(-z^2/a_{ho}^2)}$, compare Eq.(\ref{mu_profile}):
\begin{eqnarray}
\label{integ}
\Delta \mu^{LDA}_{SI} =  \int d z \, \phi_0^2(z)  \Delta \mu_{3D}(n_{\perp}\psi_0^2(z)).
\end{eqnarray}
Please note, that we depart here from dimensionless quantities and  $\Delta\mu_{3D}$ and similarly the LDA expression, $\Delta \mu^{LDA}_{SI}$ are dimensional quantities (SI units for instance). In order to compare it with  $\mu^{semi}_{LHY}$, given by Eq.~(\ref{mu_semi}), one has to divide it by $\hbar \omega_z \frac{4\pi a}{a_{ho}}$ i.e. we define  $ \mu^{LDA} = \frac{a_{ho}}{\hbar \omega_z 4\pi a}  \Delta \mu^{LDA}_{SI} $, which after integration, Eq.(\ref{integ}), reads:
\begin{equation}\label{mu_LDA}
 \mu^{LDA}(y) = y^{3/2}\frac{4 }{3\pi^{11/4}}  \sqrt{\frac{2}{5}}.
\end{equation}
By inspection of Eq.~(\ref{mSR}) and (\ref{dnSR}) one can find that 
$F_{semi}(y) \simeq G_{semi}(y) \simeq 1$  if $y \gg 1$.
Due to this fact, comparing Eqs.~(\ref{mu_semi_main}) and (\ref{mu_LDA}),
we find: 
\begin{eqnarray*}
\mu^{LDA}(y) \simeq \mu^{semi}_{LHY}(y)
\end{eqnarray*}
for $y \gg 1$.

\begin{figure}[htb]
    \centering
    \includegraphics[width=0.99\linewidth]{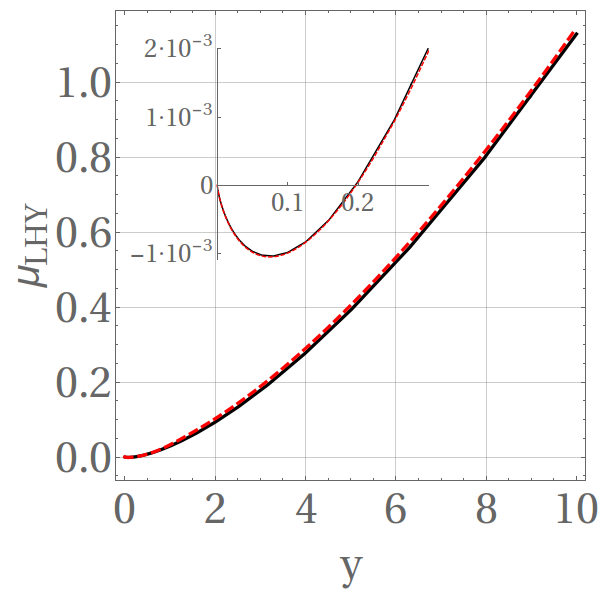}
    \caption{LHY contribution to the chemical potential, $\mu_{LHY}$, as a function of $y=\frac{gn_{\perp}}{a_{ho}\hbar\omega_z}$. Black line indicates numerical results for a harmonic confinement while the red one is the empirical fit. Note a very good agreement.}
    \label{fig:miu_fit}
\end{figure}

In the above we found analytic formulas that correctly reproduce
the numerical result in the limit $y \ll 1$ and $y \gg 1$.
To  reconstruct  the numerical results in the entire range of variations of the parameter $y$ we introduce the following empirical formula allowing for interpolation between the two above mentioned regions:
\begin{eqnarray} \label{fit}
    \mu_{LHY}(y)&=&[\mu_{LHY}^{pert}(y)+Ay^2\log{(By)}]e^{-py^2}+\nonumber\\
    & &+(1-e^{-py^2})\mu_{LHY}^{semi}(y),
\end{eqnarray}
where  $A,B,p$ are fitted parameters $A=-0.0109$, $B=1.56$ and $p=1.93$. The formula Eq.~(\ref{fit}) gently switches between  perturbative and semiclassical expression. The perturbative result, $y \ll 1$, is `enriched' by the term $\propto y^2\log{(By)}$. This modification is  a smart guess for the next order of the perturbation term. The above fit is compared to the full numerical result in Fig.~(\ref{fig:miu_fit}). We believe that this universal smooth expression can be very useful in analysis of experimental data.

\section{Summary and final remarks}

In this paper we find the beyond  mean-field contribution to the chemical potential, the so called Lee-Huang-Yang term of a two-component quantum droplet which is squeezed by a 2D harmonic potential of frequency $\omega_z$. No local density approximation is assumed in the confined direction.
The Lee-Huang-Yang contribution originates in quantum fluctuations of Bogoliubov vacuum and is proportional to sum the of quantum depletion, and  the renormalized anomalous density.

The approach is based on the Modified Gapless Hartree-Fock-Bogoliubov method which utilizes generalized Gross-Pitaevskii equation, self-consistently coupled to Bogoliubov equations. To account for a phonon-like excitations  and avoid a nonphysical gap in the spectrum we modify Bogoliubov equations by introducing a chemical potential $\mu_0$ being a solution of the zero-energy mode eigenproblem. The quantum LHY term has to be found numerically. The high energy contribution is hardly accessible because of finite spatial grid used in numerical calculations and  a singular contribution, to be treated with  care. We follow the renormalization procedure introduced in \cite{Pierwsza_praca}. The semiclassical  method is utilized to account for regularized high energy terms. 

We focus here on the universal regime, when all low energy terms $g\delta n$, $g m^R$, $\delta g n$  are much smaller than $\hbar \omega_z$. In such a situation there are only two relevant energy scales - the single component mean-field energy, $gn$, and one-particle  energy, $\hbar \omega_z$, separating ground and the first excited state of  external potential.  Physics of the problem  depends then on one universal parameter, $y=gn/\hbar \omega_z$ only. We study the Lee-Huang-Yang chemical potential in the whole extend of possible values of $y$, i.e. our results are valid in the entire range of  geometric configurations, from 3D oblate shape $\hbar \omega_z < gn$, to 2D systems where $\hbar \omega_z \gg gn$.

Results presented here contribute to the longstanding issue of the ground state energy of a weakly interacting Bose system. The Lee-Huang-Yang result is generalized to the case of nonuniform systems. This allow for a theoretical description of quantum droplets not assuming the local density approximation. We focus on  the universal regime. The obtained expressions  cover the entire range of geometries form quasi-2D to 3D settings. Of particular importance  are low dimensional systems. In such geometric arrangements  the three-body losses should be suppressed. Long-lived droplets will make possible many experiments were not only static droplets, but also dynamical situations are  subject to observation. 

\begin{acknowledgments}
This research was funded by the (Polish) National 
Science Centre Grant No. 2017/25/B/ST2/01943.
\end{acknowledgments}


\appendix


\section{Harmonic confinement - semiclassical approach} \label{app:har}

Here we  to calculate a high energy contribution to  $\mu_{LHY}$ using semiclassical approximation. From Eq.~(\ref{BogHar}) we find that the quasiparticle energies are: 
\begin{eqnarray}
&&\varepsilon =\sqrt{ \left(  \frac{k^2}{2} + \frac{1}{2} z^2 -\frac{1}{2}  \right)
\left(   \frac{k^2}{2} + \frac{1}{2} z^2 - \frac{1}{2}  + 2 y \phi_0^2(z) \right)
   }, \nonumber \\
\end{eqnarray}
In the above we notice that if $z=0$ and $k^2 < 1$ the quasiparticle energy 
is imaginary. That is due to the presence of the zero-point energy of the harmonic oscillator,
i.e. the $\frac{1}{2} $ term. This is `purely' quantum contribution which is not consistent with 
the semiclassical approximation and should be neglected: 
\begin{eqnarray}
&&\varepsilon(k,z) = \frac{1}{2} \sqrt{ \left(  k^2 +  z^2   \right)
\left(   k^2 +  z^2   + 4 y \phi_0^2(z) \right)}. 
\end{eqnarray}

The low energy region is defined as $ 0 \leq k_\perp \leq k_c$ and 
$\varepsilon_{k_\perp,\nu} < \frac{k_\perp^2}{2} + \frac{k_c^2}{2} $,
where obviously, $k^2=k_z^2+k_\perp^2$.
In the semiclassical approximation a value of the  $z$-component of momentum, $k_{z,c}$,
which separates low and high energy regions can be found from:
\begin{equation}
2 \varepsilon(k_\perp,k_{z,c}(z)) = k_\perp^2 + k_c^2,
\end{equation}
which gives:
\begin{eqnarray}\label{app111}
&&k_{z,c}^2(z,k_\perp) = 
\sqrt{  \left(k_\perp^2 + k_c^2\right)^2 + \left( 2 y \phi_0^2(z) \right)^2  } \nonumber \\
&&- \left(k_\perp^2 + z^2 + 2y \phi_0^2(z)\right), 
\end{eqnarray}
or $k_{z,c}=0$, if the above value is negative.
The high energy part of anomalous density is:
\begin{eqnarray}
&& m_{H}(z) = - 
 \frac{ y \phi_0^2(z)}{(2\pi)^3} \int_H  \mbox{d} \K \frac{e^{i\K_\perp \dx_\perp + i k_z \Delta z}}{2\varepsilon(k,z) }.
\end{eqnarray}
Now we observe that 
$\frac{1}{\varepsilon(k,z)} = \frac{1}{k_\perp^2 + k_z^2}  +\left(\frac{1}{\varepsilon(k,z)}
-\frac{1}{k_\perp^2 + k_z^2}\right)$. Using this equality we get:
\begin{eqnarray} \label{mH}
&& m_{H}(z) \simeq - \frac{y \phi_0^2(z) }{4\pi |\dx|} \\
&&   +m^0_{H}(z) +  \frac{y \phi_0^2(z)}{8\pi^2} k_c \left( \pi + \log 4   \right) h(z,y,k_c), \nonumber
\end{eqnarray}
This way the singular (in the limit $|\Delta \x| \to 0$) contribution  to the
anomalous density was found, $\sim 1/\Delta \x$ . The remaining two terms are regular. 
In particular we introduced:
\begin{eqnarray}
m^0_{H}(z) =- 
 \frac{y \phi_0^2(z) }{(2\pi)^3} \int_H   \mbox{d} \K  \, \left( \frac{1}{2\varepsilon(k,z) } - \frac{1}{k_\perp^2 + k_z^2} \right). 
\end{eqnarray}
and  function $h(z,y,k_c)$:
\begin{equation}\label{def:h}
h(z,y,k_c) = 
 \frac{1}{k_c \pi \left( \pi + \log 4  \right) }\int_L \mbox{d} \K\  \frac{1}{k_\perp^2 + k_z^2}.  
\end{equation}
Here we split the integration region $\int  \mbox{d} \K =  \int_L \mbox{d} \K + \int_H \mbox{d} \K$, where $\int_L \mbox{d} \K = \int_{k_\perp < k_c} \mbox{d} \K_\perp  
\int_{- k_{z,c}(z)}^{k_{z,c}(z)} \mbox{d} k_z $.

Utilizing explicit expressions we can bring  $m^0_{H}(z)$ to the form: 
\begin{equation}\label{mRsemW}
m^0_{H}(z) = 
 \frac{\left( y \phi_0^2(z) \right)^{3/2} }{\pi^2}   g (z,y,k_c),
\end{equation}
Where function $g(z,y,k_c)$ is defined as follows:
\begin{eqnarray}
&& g(z,y,k_c) = \frac{1}{4}
 \int_0^{\tilde x_c} \mbox{d} \tilde x  \int_{ \tilde k_{z,c}}^\infty \mbox{d} \tilde k_z \,
 t( \tilde x_0,\tilde k_z)  \nonumber \\ 
&& + \frac{1}{4} \int_{\tilde x_c}^\infty \mbox{d} \tilde x  \int_{0}^\infty \mbox{d} \tilde k_z  \,  
t( \tilde x_0,\tilde x, \tilde k_z).
\end{eqnarray}
We introduced scaled variables:   $\tilde x =\tilde k_{\perp}^2/2y\phi_0^2$, $ \tilde x_0 =z^2/2y \phi_0^2(z)  $,  $\tilde x_c = k_c^2/2y \phi_0^2(z) $, $\tilde k_{z,c}=\tilde k_{z,c}/\sqrt{2y \phi_0^2(z)}$, and $\tilde k_z=\tilde k_z/\sqrt{2y \phi_0^2(z)}$.

In the above the function  $t(\tilde x_0, \tilde x, \tilde k_z)$ is defined:
\begin{equation*}
    t(\tilde x_0, \tilde x, \tilde k_z) =  \frac{1}{\tilde x + \tilde k_z^2/2} 
-   \frac{1}{\sqrt{\left(\tilde x + \tilde k_z^2/2 + \tilde x_0 +1\right)^2 - 1}} .
\end{equation*}
Summarizing the above discussion, we found that regular part of the high
energy contribution to the anomalous regularized density, $m^R_{H}(z)$, is:
\begin{eqnarray}
&& m^R_{H}(z) =  \frac{y \phi_0^2(z)}{8\pi^2} k_c \left( \pi + \log 4   \right) h(z,y,k_c) \nonumber  \\
&& + \frac{\left( y \phi_0^2(z) \right)^{3/2} }{\pi^2}   g (z,y,k_c).
\end{eqnarray}
This high-energy component to the regularized anomalous density has to be supplemented by the low energy 
 contribution, $m_L(z)$. This should be calculated directly from the definition provided that numerical solutions of the Bogoliubov equations are found. \\

Instead, we can extend the semiclassical calculations to the entire range of excitations energies. These procedure is not legitimate in the entire range of variation of $y$. On the the hand if $y \geq 1 $ the harmonic confinement is weak and $\hbar \omega_z \ll gn $, therefore  excitation spectrum is dominated by a spectrum of free particle. In such a situation semiclassical approach can be extended over entire energy range. Semiclassical estimation of quantum depletion and anomalous density can be obtained without solving numerically Bogoliubov equations. It might be illuminating though to find  semiclassical anomalous density and quantum depletion and compare to  rigorous results. 

Semiclassical expression for a renormalized anomalous density is  analogical to Eq.(\ref{mH}). The singular  term there should be omitted (renormalization) and the cut-off dependent term vanishes. Moreover, the integration in expression giving $m^0_{H}$, Eq.(\ref{mRsemW}) has to be extended over whole momenta range. It is convenient therefore, to introduce the function $g_{semi}(\tilde x_0)= g_{semi}(z,y)$:  
\begin{eqnarray}
&& g_{semi}(\tilde x_0)
= \frac{1}{4} \int_{0}^\infty \mbox{d} \tilde x  \int_{0}^\infty \mbox{d} \tilde k_z  \, t(\tilde x_0, \tilde x, \tilde k_z).
\end{eqnarray}

Renormalized anomalous density calculated semiclassicaly  in the entire range of energies  is given by 
\begin{eqnarray} \nonumber
&& m^R_{semi}(z,y) = -
 \frac{ y \phi_0^2(z) }{(2\pi)^3}  \int \mbox{d} \K  \left( \frac{1}{2\varepsilon(k,z) } - \frac{1}{k^2} \right) 
\\ \label{mRsemPP}
&&
=  \frac{1}{\pi^2}  \left( y \phi_0^2(z) \right)^{3/2}   g_{semi} \left(\frac{z^2}{2y \phi_0^2(z)} \right).
\end{eqnarray}\\

Now we turn to calculation of the quantum depletion.
The high-energy part of quantum depletion obtained within the  semiclassical approximation takes the form
\begin{eqnarray} \label{n_H}
&&\delta n_H(z,y,k_c) = \frac{\left( y \phi_0^2(z) \right)^{3/2}}{3 \pi^2} 
 f(z,y,k_c) \\
&& \equiv \frac{1}{2} \int_H \frac{\mbox{d} \K}{(2\pi)^3} 
\left( \frac{ (k_\perp^2 + k_z^2) + (z^2-1) + 2 y \phi_0^2(z) }{ 2\varepsilon(k,z) } - 1 \right), \nonumber\label{dnH1} \nonumber
\end{eqnarray}
where we introduced function $f(\tilde x_0, \tilde x_c) = f(z,y,k_c)$:
\begin{eqnarray}
&&f(\tilde x_0, \tilde x_c)
= \frac{3}{4}
 \int_0^{\tilde x_c} \mbox{d} \tilde x  \int_{ \tilde k_{z,c}}^\infty \mbox{d} 
 \tilde k_z 
 s(\tilde x_0, \tilde k_z) \nonumber \\
&& + \frac{3}{4} \int_{\tilde x_c}^\infty \mbox{d} \tilde x  \int_{0}^\infty \mbox{d} \tilde k_z  s(\tilde x_0, \tilde k_z), 
\end{eqnarray}
and  $s(\tilde x_0, \tilde k_z)=s(z,y,k_c)$: 
\begin{eqnarray}
&&s(\tilde x_0, \tilde k_z)
\frac{ \tilde x + \frac{ \tilde k_z^2}{2} + \tilde x_0 + 1 }{
\sqrt{ \left( \tilde x + \frac{\tilde k_z^2}{2} + \tilde x_0 +1  \right)^2 - 1}} - 1.
\end{eqnarray}
Again, the high-energy contribution to the quantum depletion has to be supplemented
by the low-energy term $\delta n_L$ which depends on solutions of the Bogoliubov equations and should be found directly from the definition by numerical computations.

Similarly as in the case of anomalous regularized density, we can extend integration
over the entire range of energies and obtain the semiclassical expression: 
\begin{equation}\label{ndsemPP}
\delta n_{semi}(z,y) =  \frac{1}{3\pi^2}  \left( y \phi_0^2(z) \right)^{3/2}   f_{semi} \left(\frac{z^2}{2y \phi_0^2(z)} \right),
\end{equation}
where we introduced $f_{semi}(\tilde x_0) = f_{semi}(z,y)$:
\begin{eqnarray}
 f_{semi}(\tilde x_0) = \frac{3}{2\sqrt{2} }   \int_{\tilde x_0}^\infty \ d \tilde x 
 \sqrt{\tilde x- \tilde x_0} 
\left( \frac{ \tilde x+ 1 }{ \sqrt{ \tilde x(\tilde x + 2) } } - 1 \right). \nonumber 
\\
\end{eqnarray}

\begin{widetext}
While calculating $\mu_{LHY}$ we arrive at a number of integrals for which we introduce a shorthand notation: 
\begin{equation}\label{mSHR}
\int dz \, m_{H}^R(z) |\phi_0(z)|^2
= 
\frac{y^{3/2}}{\pi^2} \int d z \,     \phi_0^5(z)  
 g \left(  \frac{z^2}{2y \phi_0^2(z)} ,   \frac{k_c^2}{2y \phi_0^2(z)}  \right)
\equiv
\frac{y^{3/2}}{\pi^{11/4}}  \sqrt{\frac{2}{5}} G(y,k_c).
\end{equation} 

\begin{equation} \label{dnSHR}
\int dz \, \delta n_{H}(z) |\phi_0(z)|^2
\equiv
\frac{y^{3/2}}{3\pi^2} \int d z \,     \phi_0^5(z)  
f \left(  \frac{z^2}{2y \phi_0^2(z)} ,   \frac{k_c^2}{2y \phi_0^2(z)}  \right) 
\equiv
\frac{y^{3/2}}{3\pi^{11/4}}  \sqrt{\frac{2}{5}}  F(y,k_c). 
\end{equation}
\begin{equation}\label{HS} 
\int \mbox{d} z \, y k_c  \frac{ \pi + \log 4  }{8 \pi^2}  
   \phi_0^4(z)   h(z,y,k_c)  \equiv  y \, k_c  \frac{ \pi + \log 4  }{8 \sqrt{2} \pi^{5/2}} H(y).
\end{equation}
\end{widetext}

These integrals, while brought together, give a high-energy contribution, 
$\mu^H_{LHY}$, to the chemical
potential of a symmetric two-component quantum droplet squeezed in $z$-direction by an external harmonic potential:
\begin{eqnarray} \label{delta_high}
&& \mu^H_{LHY} =  \frac{y^{3/2}}{3\pi^{11/4}}  \sqrt{\frac{2}{5}}\left(3 G(y,k_c)+   F(y,k_c) \right) \nonumber \\
&&  + y \, k_c  \frac{ \pi + \log 4  }{8 \sqrt{2} \pi^{5/2}} H(y).
\end{eqnarray}

If the semiclassical approach is extended over the whole range   of the energy spectrum, the  following integrals are essential:
\begin{eqnarray} \label{mSR}
\int dz \, m_{semi}^R(z) |\phi_0(z)|^2 &\equiv& \frac{y^{3/2}}{\pi^{11/4}}  \sqrt{\frac{2}{5}} G_{semi}(y), 
\end{eqnarray}
\begin{eqnarray} \label{dnSR}
\int dz \, \delta n_{semi}(z) |\phi_0(z)|^2 &\equiv& \frac{y^{3/2}}{3\pi^{11/4}}  \sqrt{\frac{2}{5}} F_{semi}(y). 
\end{eqnarray}
A semiclassical expression for the chemical potential   resulting from  quantum fluctuations is:
\begin{equation}\label{mu_semi}
  \mu^{semi}_{LHY}(y) = \frac{y^{3/2}}{3\pi^{11/4}} \sqrt{\frac{2}{5}}\left( 3 G_{semi}(y) + F_{semi}(y) \right).
\end{equation}

We want to stress that all integrals involved in the final results are regular and free of any singularities. Their numerical evaluation does not present any technical problems. Their evaluation  directly from  definitions is practically impossible. Finite spatial greed introduces a  cut-off in high-momentum space and, because of a singularity at high momenta, gives uncontrolled results.

\section{Harmonic confinement - perturbative approach} \label{har2}

In this section we consider a special case of a very strong harmonic confinement
in $z$-direction.  Because  oscillator excitation energy,   $\hbar \omega_z$, is large as compared to all other energy scales,   the chemical potential in particular, we will use perturbative solutions of  Bogoliubov equations. We assume that Bogoliubov modes are very similar to the oscillator eigenstates and are  only slightly perturbed by the mean-field interaction.    A ratio of the mean-field interaction energy to the excitation energy $y=  g n_\perp /( a_{ho}\hbar \omega_z) \ll 1$ is the small parameter.
We introduced here the 2D atomic density $n_\perp$ in the $x-y$ plane.
Perturbative calculations allow for  a comparison of our results to those of  \cite{Petrov16}, obtained  using another method.\\

We start from rewriting the Bogoliubov equations Eq.~(\ref{BogHar}) and introducing  $f^\pm_\nu = u_\nu \mp v_\nu $:
\begin{eqnarray}\label{con1}
&&\left(  x + H_z  2 y \phi_0^2(z)\right) f^-_{x,\nu} = \varepsilon_{x,\nu} f^+_{x,\nu}, \\
\label{con2}
&&\left(  x + H_z  \right) f^+_{x,\nu} = \varepsilon_{x,\nu} f^-_{x,\nu},
\end{eqnarray}
where $H_z = - 1/2 \partial_z^2 + 1/2 z^2 - 1/2$ and $x = \frac{k_\perp^2}{2}$.
\\

Next we find that: 
\begin{eqnarray} \label{Bogg1}
&& \left(  x + H_z  + 2 y \phi_0^2(z) \right) \left(  x + H_z \right) f^+_{x,\nu} = \varepsilon^2_{x,\nu} f^+_{x,\nu},
\\ \nonumber
\\  \label{Bogg2}
&& \left(  x + H_z   \right)  \left(  x + H_z  + 2 y \phi_0^2(z) \right)
f^-_{x,\nu} = \varepsilon_{x,\nu}^2 f^-_{x,\nu}.
\end{eqnarray}
By bringing   Bogoliubov equations Eqs.~(\ref{con1}), (\ref{con2}) to the form 
above, Eqs.~(\ref{Bogg1}), (\ref{Bogg2}), we effectively   'squared' them.
We observe now that these squared equations are Bogoliubow equations for
the `squared' free Hamiltonian:
\begin{equation}
H_0 =  \left( x + H_z \right)^2,
\end{equation}
with  effective interactions
\begin{eqnarray} \label{h+}
H_+ =  2 y \phi_0^2(z)  \left(  x + H_z \right), \\
\label{h-}
H_- =   \left(  x + H_z \right)2 y \phi_0^2(z), 
\end{eqnarray}
so that Eqs.~(\ref{Bogg1}) and (\ref{Bogg2}) take the form
\begin{eqnarray}
(H_0 + H_+) f^+_{x,\nu} = E_{x,\nu} f^+_{x,\nu}, \\
(H_0 + H_-) f^-_{x,\nu} = E_{x,\nu} f^-_{x,\nu},
\end{eqnarray}
where $ E_{x,\nu} = \varepsilon^2_{x,\nu}. $ The zero order equation is:
\begin{equation}
H_0 f^{\pm (0)}_{x,\nu}(z) =  E_{x,\nu}^{(0)} f^{\pm (0)}_{x,\nu}(z),
\end{equation}
and has solutions: 
\begin{eqnarray}
f^{\pm (0)}_{x,\nu}(z) = f^{\pm (0)}_{x,\nu} \phi_\nu(z), \\
 E_{x,\nu}^{(0)} =(\nu + x)^2,
\end{eqnarray}
where $\phi_\nu(z)$ is the eigenstate of the harmonic oscillator normalized to unity
i.e. $\int d z  \,  |\phi_\nu(z)|^2 =1$.

Now we turn to the the first order perturbation theory. For $\nu=0$  the Bogoliubov energy is :
\begin{equation}\label{var}
 \varepsilon^2_{x,0}  = E_{x,0}^{(0)} + E_{x,0}^{(1)}  = x^2 + 2 x y c_{0,0} , 
\end{equation}
and corresponding eigenmodes  are:
\begin{eqnarray}  \label{fpm1} \nonumber
&& f^{+}_{x,0}(z) \simeq f^{+(0)}_{x,0}  \left(\phi_0(z) 
+   \sum_{\nu > 0} \frac{   2 y  x c_{\nu,0} }{ x^2 - (x+ \nu)^2 } \phi_{\nu}(z)\right), 
\\ \nonumber
&& f^-_{x,0}(z) =  f^{-(0)}_{x,0} \left(\phi_0(z)  +   \sum_{\nu > 0} \frac{   2 y  (x + \nu) c_{\nu,0} }{ x^2 - (x+\nu)^2 } \phi_{\nu}(z)\right),
\end{eqnarray}
where $c_{\nu,\nu'} = \int \mbox{d} z \,  \phi_\nu^*(z) \phi_0^2(z) \phi_{\nu'}(z)$. In particular  $c_{0,0} =  1/\sqrt{2\pi}  $.

Inserting the above into Eq.~(\ref{con2}) we obtain
\begin{eqnarray}
&& x f^{+(0)}_{x,0}\left( \phi_0(z) + \sum_{\nu > 0} \frac{ 2 y (x+\nu)  c_{\nu,0} }{ x^2 - (x+ \nu)^2 } \phi_{\nu}(z) \right) \\
&& = \varepsilon_{x,0} f^{-(0)}_{x,0} \left(   \phi_0(z)  +   \sum_{\nu > 0} \frac{   2 y  (x + \nu) c_{\nu,0} }{ x^2 - (x+\nu)^2 } \phi_{\nu}(z) \right), \nonumber
\end{eqnarray}
where $\varepsilon_{x,0} $ is given by Eq.~(\ref{var}).
From the above we find that
\begin{equation}
x f^{+(0)}_{x,0} = \varepsilon_{x,0} f^{-(0)}_{x,0}.
\end{equation}
We additionally have the normalization condition which in the first order of perturbation reads
\begin{equation}\label{norm:cond}
f^{+(0)}_{x,0} f^{-(0)}_{x,0}  = 1.
\end{equation}
This completes the first order calculation for $\nu = 0$ component. We
found:
\begin{eqnarray}
\label{e0}
\varepsilon_{x,0} = \sqrt{ x^2 + 2 x y c_{0,0} }, \\
\label{f0}
f^{\pm(0)}_{x,0} =  \left(  \frac{\varepsilon_{x,0}}{x} \right)^{\pm 1/2},  
\end{eqnarray}
which together with Eq.~(\ref{fpm1}) give  explicit expressions for $f^\pm_{x,0}(z)$. Thus $\nu=0$ mode contribution to $\mu_{LHY}$   is given by the term $ \sim \bigg(m(z) + \delta n(z)\bigg)_{\nu=0} \equiv -\frac{1}{4\pi}\int {\rm d}x {\cal F}_{x,0}(z)$.
The function ${\cal F}_{x,0}(z)$ is, up to linear in $y$ terms, given by 
\begin{eqnarray}
&& {\cal F}_{x,0}(z) = f^+_{x,0}(z)f^-_{x,0}(z)   - (f^-_{x,0}(z))^2  \nonumber  \\
&&\simeq  \phi_0^2(z) \left( 1 -   (f^-_{x,0,0})^2 \right) + 
\\
&& + 2 y  \phi_0(z) 
  \sum_{\nu > 0} \frac{1}{\nu}  \left(    \frac{x}{\varepsilon_{x,0}} \frac{2(x+\nu)}{ 2 x  + \nu }  -  1  \right) c_{\nu,0} \phi_{\nu}(z) \nonumber.
\end{eqnarray}
However in the first order of perturbation we approximate
$\frac{x}{\varepsilon_{x,0}} \frac{2(x+\nu)}{ 2 x  + \nu } 
\rightarrow \frac{2(x+\nu)}{ 2 x  + \nu }  $ 	
to get:
\begin{eqnarray}
&& {\cal F}_{x,0}(z)  
\simeq \phi_0^2(z) \left( 1 -   (f^-_{x,0})^2 \right)
\\
&&
+ 2 y  \phi_0(z) 
  \sum_{\nu > 0} \frac{1}{ 2 x  + \nu }  
c_{\nu,0} \phi_{\nu}(z)
\end{eqnarray}

Analogically we obtain $\nu > 0 $  contributions -- Bogoliubov energies:
\begin{equation}\label{varnu}
\varepsilon^2_{x,\nu} = (x + \nu)^2 +2 y (x + \nu) c_{\nu,\nu}  
\end{equation}
and wavefunctions:
\begin{eqnarray}
&& f^+_{x,\nu}(z) = 
f^{+(0)}_{x,\nu} \left( \phi_\nu(z) +  \sum_{\nu' \neq \nu} \frac{ 2 y  (x+ \nu)  c_{\nu',\nu} \phi_{\nu'}(z)}{ (x+\nu)^2 - (x+ \nu')^2 } \right) , \nonumber \\
&& f^-_{x,\nu} (z) =    
f^{-(0)}_{x,\nu} \left( \phi_\nu(z)  + \sum_{\nu' \neq \nu} \frac{ 2 y  (x + \nu') c_{\nu',\nu} \phi_{\nu'}(z)}{ (x+\nu)^2 - (x+ \nu')^2 }   \right). \nonumber 
\end{eqnarray}
By inserting the above into Eq.~(\ref{con2}) we arrive at 
\begin{equation}
  f^{+(0)}_{x,\nu}  (x+\nu)  = \varepsilon_{x,\nu}   f^{-(0)}_{x,\nu}.
\end{equation}
Using the normalization condition,  $ f^{+(0)}_{x,\nu}f^{-(0)}_{x,\nu} =1  $, the 
amplitudes of excited modes are:
\begin{equation}
  f^{\pm(0)}_{x,\nu}  = \left( \frac{\varepsilon_{x,\nu}}{x+\nu} \right)^{\pm 1/2},  
\end{equation}
what allows to find an integrand ${\cal F}_{x,\nu}= f^+_{x,\nu}(z)f^-_{x,\nu}(z)   - (f^-_{x,\nu}(z))^2$ for  $\nu>0$:
\begin{eqnarray}
&&{\cal F}_{x,\nu}= \frac{y c_{\nu,\nu}}{x + \nu}   \phi_\nu^2(z)
+ 
\phi_\nu(z) \sum_{\nu' \neq \nu} \frac{   2 y c_{\nu',\nu} }{ 2x + \nu + \nu'} \phi_{\nu'}(z) \nonumber
\\
&&= \phi_\nu(z) \sum_{\nu' } \frac{   2 y c_{\nu',\nu} }{ 2x + \nu + \nu'} \phi_{\nu'}(z).
\end{eqnarray}

From the above we obtain
\begin{eqnarray} \label{mz}
&& m_L(z) + \delta n_L(z) =
\\
&&
-  \frac{1}{2\pi} \int_0^{x_c} \mbox{d} x \,
\sum_{ 0< \nu \leq x_c}   y \phi_\nu(z) \sum_{\nu' } \frac{ c_{\nu,\nu'}}{ 2x + \nu + \nu'} 
\phi_{\nu'}(z). \nonumber 
\end{eqnarray}
In Eq.(\ref{mz}) we account only for low energy excitation of the system. We set the upper limit of integration as well as summation finite.  We remind that we introduced notation $\frac{k_\perp^2}{2} = x$ and $\frac{k_c^2}{  2} = x_c$. In sec. \ref{harm2d}  the low energy sector is defined  as $\varepsilon_{k_\perp,\nu} \leq  \frac{k_\perp^2}{2} + \frac{k_c^2}{2} $ and $k_\perp \leq k_c$.  
Now, we must set the upper limit of summation in Eq.(\ref{mz}). 

From Eq.~(\ref{varnu}) we have
\begin{equation}
\varepsilon_{x,\nu} - (x + \nu) \simeq  y  \int dz \, \phi_\nu^2(z) \phi_0^2(z)
\propto \frac{y}{\sqrt{\nu}},
\end{equation}
i.e. for large $\nu$ the energy $\varepsilon_{x,\nu}$ can be approximated by $\varepsilon_{x,\nu} \simeq (x + \nu) $. In the view of this equality one can clearly see that the condition $\varepsilon_{k_\perp,\nu} \leq  \frac{k_\perp^2}{2} + \frac{k_c^2}{2} = x+ x_c $ will be met for $\nu \leq x_c$. This justifies why the upper limit of summation is  $\nu=[x_c]$, where $[x_c]$ is integer part of $x_c$.
On the other hand summation over $\nu'$ -- the intermediate states, is extended up to infinity.

Writing explicitly contribution form  $\nu=0$ mode, Eq.~(\ref{mz}) becomes  
\begin{eqnarray}
&& m_L(z) + \delta n_L(z) =  -  \frac{1}{4\pi} \int_0^{x_c} \mbox{d}x \, 
\bigg(  \phi_0^2(z)( 1 -   (f_{x,0}^-)^2) \nonumber 
\\
&& 
+  y    \sum_{\nu' > 0} \phi_0(z)  \phi_{\nu'}(z) \frac{c_{\nu',0}}{ x  + \frac{\nu '}{2}} 
\nonumber
\\
&&
 + 
y  \sum_{ 0 < \nu \leq x_c}  \sum_{\nu' }  \phi_\nu(z)\phi_{\nu'}(z) 
\frac{ c_{\nu,\nu'} }{ x + \frac{\nu + \nu'}{2}}  \bigg).
\end{eqnarray}
Contribution to the chemical potential from the LHY energy is $\mu^{sng}_{LHY} = \int {\rm d}z \phi_0(z)^2(m_L(z) + \delta n_L(z))$,  therefore:
\begin{eqnarray}
&& -4 \pi \mu^{sng}_{LHY} =  \int_0^{x_c} \mbox{d} x \left( (  c_{0,0} ( 1 -   (f_{x,0}^-)^2)   +y   
     \sum_{\nu' > 0}  \frac{c_{\nu',0}^2}{ x  + \frac{\nu '}{2}}\right.\nonumber  
\\
&&
\left. 
 + y
 \sum_{ x_c \geq \nu > 0 }  \sum_{\nu' }  
\frac{ c_{\nu,\nu'}^2 }{ x + \frac{\nu + \nu'}{2}}  \right).
\end{eqnarray}

$\mu^{sng}_{LHY}$ contains a singular contribution. LHY chemical potential  $\mu_{LHY}$ 
involves not anomalous, $m_L(z)$, but regularized anomalous density, $m^R(z)$. As follows from  Eq.~(\ref{hmr}), the regularization in limit of 
$k_c \rightarrow \infty$  amounts to addition of a cut-off depending term (the final result does not depend on cut-off, however): 
$ m^R(z) = m_L(z) + y \phi_0^2(z)  \frac{k_c}{8\pi^2}  (\pi + \log 4) $. 
Therefore chemical potential $\mu_{LHY}$ is: 
\begin{equation}
\label{reg}
    \mu_{LHY} =  \left( \mu^{sng}_{LHY} + y c_{0,0} \frac{k_c}{8\pi^2}  (\pi + \log 4)\right).
\end{equation}
The second term in Eq.(\ref{reg}), can be  written in the integral form:
\begin{eqnarray*}
&&y c_{0,0} \frac{k_c}{8\pi^2}  (\pi + \log 4)=\nonumber \\
&&\frac{y c_{0,0}}{4\pi} \int_0^{x_c} \mbox{d}x  \frac{1}{\pi \sqrt{2}} \int_0^{x_c} \mbox{d}\nu_z \,  \frac{1}{\sqrt{\nu_z}}  \frac{1}{x + \nu_z}.
\end{eqnarray*}
As a result we have:
\begin{eqnarray}
&&\mu_{LHY } = \\
&&
=  -  \frac{1}{4\pi} \int_0^{x_c} \mbox{d}x \,
\bigg(  c_{0,0} ( 1 -   (f^-_{x,0,0})^2)
+  y    \sum_{\nu' > 0} \frac{c_{\nu',0}^2}{ x  + \frac{\nu '}{2}} \nonumber
\\
&&
+ 
y  \sum_{ x_c \geq \nu > 0}  \sum_{\nu' } 
\frac{ c_{\nu,\nu'}^2 }{ x + \frac{\nu + \nu'}{2}}  
 - y \frac{1}{\pi}  \int_0^{x_c} d\nu_z \, \frac{1}{\sqrt{2 \nu_z}} \frac{c_{0,0}  }{x +  \nu_z} 
\bigg) \nonumber
\end{eqnarray}
Using $(f^{-(0)}_{x,0})^2 = \left( \frac{x}{\varepsilon_{x,0}}\right) $, Eq.(\ref{e0}), 
and $\varepsilon_{x,0} = \sqrt{ x^2 + 2 x y c_{0,0} }$, Eq.(\ref{f0}), we get:
\begin{eqnarray}
&&\int_0^{x_c} \mbox{d}x \,  c_{0,0} ( 1 -   (f^-_{x,0,0})^2) =\\
&&\int_0^{x_c} \mbox{d}x \,  \frac{2  y c_{0,0}^2 x}{(x + \varepsilon_{x,0}) \varepsilon_{x,0}} =   y c_{0,0}^2 \left(  - 1 + \log  \left( \frac{2x_c}{y c_{0,0}}  \right)   \right). \nonumber
\end{eqnarray}
So finally we have (using $c_{0,0} = 1/\sqrt{2\pi}$)
\begin{widetext}
\begin{eqnarray} \label{semicl}
&&   \mu_{LHY }(y)=-   \frac{y}{8\pi^2}\left[   - 1 + \log 2   - \log y + \frac{1}{2} \log  (2\pi) 
+
 2\pi \int_0^{x_c} \mbox{d}x \,  \sum_{\nu' > 0} \frac{c_{\nu',0}^2}{ x  + \frac{\nu '}{2}} \right.
\\
&& 
\left. +   \log x_c
+
 \int_0^{x_c} \mbox{d}x \,
\left(
   2\pi  \sum_{ x_c \geq \nu > 0}  \sum_{\nu' } 
\frac{ c_{\nu,\nu'}^2 }{ x + \frac{\nu + \nu'}{2}}  
 -  \frac{1}{\sqrt{\pi}}  \int_0^{x_c} \mbox{d}\nu_z \, \frac{1}{ \sqrt{\nu_z}} \frac{1  }{x +  \nu_z} 
\right) \right]. \nonumber
\end{eqnarray}
\end{widetext}
In the above one can identify a term proportional to $y \log y$ and terms proportional to
to $y$. Therefore the final formula can be simplified and written in the form:
\begin{equation}\label{wynikP}
 \mu^{pert}_{LHY} \equiv \mu_{LHY}(y) = \frac{y}{8\pi^2} \log \left(  y \frac{C_{2d}^h \sqrt{e}}{4\pi}   \right). 
\end{equation}
This is the perturbative expression for the LHY contribution to the chemical potential, valid 
in the limit of a very strong confinement in $z$-direction, $y \ll 1$. 
Summations and integrations in Eq.~(\ref{semicl}) can be completed in quite tedious   calculations 
in the limit of $x_c \rightarrow \infty$
giving the value   $C_{2d}^h \simeq 28.69 $ which is in agreement with \cite{Petrov16}.

\bibliography{main}
\bibliographystyle{apsrev4-1}

\end{document}